\documentclass{rsifpublic}
\usepackage{graphicx}
\usepackage{amssymb}
\usepackage{natbib}
\bibpunct{(}{)}{;}{a}{}{,}
\begin{document}
\runningheads{Hennig, Archilla, Romero}{Enzymes and bubbles in
{DNA}}

\begin{topmatter}{
\title{
Modeling the thermal evolution of enzyme-created bubbles in {DNA}}

\author{D. Hennig\affil{1}, J.
F. R. Archilla\affil{2}\comma\corrauth\ and J. M. Romero\affil{2}
}
\address{
\affilnum{1} Freie Universit\"{a}t Berlin, Fachbereich Physik,
Institut f\"{u}r Theoretische Physik,\\ Arnimallee 14, 14195
Berlin, Germany\\ \affilnum{2} Nonlinear Physics Group, University
of Sevilla,
 Dep.  F\'{\i}sica Aplicada I, ETSI Inform\'atica,
  Avda Reina Mercedes s/n, 41012 Sevilla, Spain
  }

\date{November 29, 2004}
\begin{abstract}
The formation of bubbles in nucleic acids (NAs) are fundamental in
many biological processes such as DNA replication, recombination,
telomeres formation, nucleotide excision repair, as well as RNA
transcription and splicing. These precesses are carried out by
assembled complexes with enzymes that separate selected regions of
NAs. Within the frame of a nonlinear dynamics approach we model
the structure of the DNA duplex by a nonlinear network of coupled
oscillators. We show that in fact from certain local structural
distortions there originate oscillating localized patterns, that
is radial and torsional breathers, which are associated with
localized H-bond deformations, being reminiscent of the
replication bubble. We further study the temperature dependence of
these oscillating bubbles. To this aim the underlying nonlinear
oscillator network of the DNA duplex is brought in contact with a
heat bath using the Nos$\rm{\acute{e}}$-Hoover-method. Special
attention is paid to the stability of the oscillating bubbles
under the imposed thermal perturbations. It is demonstrated that
the radial and torsional breathers, sustain the impact of thermal
perturbations even at temperatures as high as room temperature.
Generally, for nonzero temperature the H-bond breathers move
coherently along the double chain whereas at $T=0$ standing radial
and torsional breathers result.
\end{abstract}

 }\end{topmatter}
\corraddr{archilla@us.es}

\section{Introduction}
The vital life processes of nucleic acids (NAs) have become
increasingly addressed and intensively studied over the last years
by biologists, chemists, mathematicians and physicists. The
interest of the latter is mainly focused on the actual dynamical
processes involved in information retrieval from the NAs duplexes
associated with structural transitions to single-stranded
sequences. Consequently, the two complementary strands of double
helical DNA must be separated (or at least locally melted) and
rearranged in order to yield single-stranded sequences to be used
by other molecules, requiring the breaking up of the hydrogen
bonds in the corresponding region of the double helix.  Therefore,
the opening of a select region of the twisted (Watson-Crick) DNA
double helix, begins with a partial unwinding at an area called
the replication fork and is observed as a bubble. In general, the
unwinding of NAs is achieved by a type of enzyme belonging to the
large family  of nucleic acid helicases (see \citealt{Stryer})
(more than sixty different types).

They are ubiquitous and versatile enzymes that, in conjunction
with other components of the macromolecular machines, carry out
important biological processes such as DNA replication, DNA
recombination, nucleotide excision repair (i.e. UV damage), RNA
editing and splicing, transfer of single-stranded nucleic acid
(ssNA) to other ssNA, protein or release into solution
 \citep{Hippel01}
   and formation of T-loop telomeres
  \citep{Wu01}.
Nucleic acid helicases have several structural varieties
(monomeric, dimeric, trimeric, tetrameric and closed hexameric),
but all of them use the hydrolysis of nucleoside triphosphate
(NTP) to nucleoside diphosphate (NDP) as the preferred source of
energy. Our present study deals with a nonlinear dynamics model of
the formation of oscillating bubbles in regions of duplex DNA
acted upon by an enzyme. Utilizing models based on nonlinear
lattice dynamics to study the formation of DNA opening has been of
great interest lately
 \citep{Englander,Agarwal}.
   The existence of localized modes, such as solitons and breathers
(describing energy localization and coherent transport),  makes
the nonlinear lattice approaches  appealing.
 With view
to vibrational excitation in DNA the Peyrard--Bishop (PB)
 model~\citep{PB}
 and
its successors \citep{Dauxois,BCP,Cocco99,Barbi99,Barbithesis}
   have
been successfully applied to describe moving localized excitations
(breathers) which reproduce typical features of the DNA opening
dynamics such as the magnitude of the amplitudes and the time
scale of the oscillating 'bubble' preceding full strand
separation.

We are interested in the bubble formation process initiated by
structural deformations of selected regions of the parental DNA
duplex that serves as a template for the replication. The process
begins when the replication apparatus identifies the starting
point and then gets combined to it. The replication apparatus is
and assembled complex that forms the replication fork and opens it
directionally. During this process the helical DNA is unwound and
replicated. One of the components of that complex is the helicase,
an enzyme unable to recognize the origin of the replication {\it
per se}, and which requires the participation of specific proteins
to lead it to the initiation site \citep{Delagoutte03}.

It is assumed that this enzyme operates in the way, that a local
unwinding in combination with (rather small) stretchings of the
H-bonds in this region occurs. We aim to demonstrate that there
are indeed initial deformations that give rise to localized
vibrations, constrained to a region of DNA, matching the
properties of oscillating bubbles observed experimentally in DNA.
These oscillating bubbles with their temporarily extended but yet
unbroken H-bonds serve as the precursors to the replication
bubble. Furthermore, we are also interested whether the stable
radial and torsional breathers persist under imposed thermal
perturbations.

\section{Methods}
The nonlinear oscillator network model for the DNA double helix
used in this paper is explained in detail in
\citep{BCP,Cocco99,Agarwal,Hennig}. The equilibrium position of
each base within the duplex configuration is described in a
Cartesian coordinate system by $x_{n,i}^{(0)}$, $y_{n,i}^{(0)}$
and $z_{n,i}^{(0)}$. The index pair $(n,i)$ labels the $n$-th base
on the $i-$th strand with $i=1,2$ and  $1\leqq n\leqq N$\,, where
$N$ is the number of base pairs considered. Displacements of the
bases from their equilibrium positions are denoted by $x_{n,i}$,
$y_{n,i}$ and $z_{n,i}$. The potential energy taking into account
the interactions between the bases consists of four parts. The
potential energy of the hydrogen bond within a base pair is
modeled typically by a Morse potential
\begin{equation}
V_{h}^n=D_n \left[\,\exp\left(-\frac{\alpha}{2}\,
d_n\right)\,-1\,\right]^2\,, \label{eq:Uhyd}
\end{equation}
where the variables $d_n$ describe dynamical deviations of the
hydrogen bonds from their equilibrium lengths $d_0$ (for details
see \citealt{Hennig}). The site-dependent depth of the Morse
potential, $D_n$, depends on the  number of involved hydrogen
bonds for the two different pairings in DNA, namely the G-C and
the A-T pairs. The former pair includes three hydrogen bonds while
the latter includes only two. $\alpha^{-1}$ is a measure of the
potential--well width.

The energies of the rather strong and rigid covalent bonds between
the nucleotides $n$ and $n-1$ on the $i-th$ strand are modeled by
 harmonic potential terms
\begin{equation}
V_{c}^{n,i}=\frac{K}{2}\,l_{n,i}^2\,, \label{eq:Uback}
\end{equation}
and $l_{n,i}$ describes the deviations from the  equilibrium
distance between two adjacent bases on the same strand. $K$ is the
elasticity coefficient.

Effects of stacking, which impede that, due to the backbone
rigidity, one base slides over another \citep{Stryer} are
incorporated in the following potential terms
\begin{equation}
V_{s}^{n,i}=\frac{S}{2}\,(d_{n,i}-d_{n-1,i})^2\,.
\end{equation}
The supposedly small deformations in longitudinal direction can be
modeled by harmonic elasticity potential terms given by
\begin{equation}
V_{l}^{n,i}=\frac{C}{2}\,(z_{n,i}-z_{n-1,i})^2\,.
\end{equation}
The kinetic energy of a nucleotide is determined by
\begin{equation}
E_{kin}^{n,i}=\frac{1}{2m}\,\left[\,\left(p_{n,i}^{(x)}\right)^2
+\left(p_{n,i}^{(y)}\right)^2
+\left(p_{n,i}^{(z)}\right)^2\right]\,,\label{eq:Ekin}
\end{equation}
where $m$ is the mass and $p_{n,i}^{(x,y,z)}$ denotes the
$(x,y,z)-$component of the momentum.

The model Hamiltonian reads then as
\begin{equation}
H=\sum_{i=1,2}\,\sum_{n=1}^{N}\,E^{n,i}\,,\label{eq:Hdna}
\end{equation}
with
\begin{equation}
E_{n,i}=E_{kin}^{n,i}+V_{h}^{n}+V_{c}^n+V_{s}^{n,i}+V_{l}^{n,i}\,,\label{eq:Eloc}
\end{equation}
and the summation in (\ref{eq:Hdna}) is performed over all
nucleotides and the two strands.

For a typical equilibrium configuration the rotation angle, by
which each base is rotated around the central axis, is given by
$\theta_0=36^\circ$, the distance between base pair planes is
$h=3.4$\,\AA\,, and the inter-base distance (the diameter of the
helix) is $d_0=20$\,\AA. For the average mass of one nucleotide we
use $M=300\,\mathrm{amu}=4.982\times10^{-25}\mathrm{kg}$.
 The arbitrary base pair sequence is
reflected in randomly distributed bi-valued H-bond coupling
strengths $D_0$ (for A-T) and $D_1=2\,D_0$ (for G-C). Note that
the ratio 1.5 is often used due to the fact that the G-C base pair
has 3 hydrogen bridges and the A-T two of them, but the energy
depends on the angles and distances of the H-bonds and on the
polarity of the bases among other factors. Quantum chemical
calculations \citep{Sponer} lead to the ratio 2 as used here.
However, the ratio 1.5 lead to qualitatively similar results.
 The aperiodic arrangement of the two different
base-pairs, coding the genetic information, renders the DNA chain
to be of {\it A-B-disorder} type. Following Barbi et al
\citeyearpar{BCP} we set $\alpha=4.45$\,\AA$^{-1}$,
$D_0=0.04\,\mathrm{eV}$, and $K=1.0\,\mathrm{eV}$\,\AA$^{-2}$. The
value of the parameter $S$ is reasonably taken as $S=2K$
\citep{Cocco99} and a credible value for $C$ is given by $C=S/10$
\citep{Agarwal}.

With the time scaled as $t\rightarrow \sqrt{D_0 \alpha^2/m}\,t$
one passes to a dimensionless formulation with quantities:
\begin{eqnarray}
 \tilde{x}_{n,i}=\alpha x_{n,i}\,, \,\,\,\,
\tilde{y}_{n,i}=\alpha y_{n,i}\,,\,\,\,\, \tilde{z}_{n,i}=\alpha
z_{n,i}
\\
 \tilde{p}_{n,i}^{(x)}=\frac{p_{n,i}^{(x)}}{\sqrt{mD_0}}\,,\,
\tilde{p}_{n,i}^{(y)}=\frac{p_{n,i}^{(y)}}{\sqrt{mD_0}} \,,\,
\tilde{p}_{n,i}^{(z)}=\frac{p_{n,i}^{(z)}}{\sqrt{mD_0}} \,,
 \\
  \tilde{D}_1=\frac{D_1}{D_0}\,,\,
  \tilde{C}=\frac{C}{\alpha^2
 D_0}\,,\,
 \tilde{K}=\frac{K}{\alpha^2 D_0}\,,\,
\tilde{S}=\frac{S}{\alpha^2 D_0}\,,
 \\
  \tilde{d}_{n}=\alpha\,
d_{n}\,,\,\,\,\,
\tilde{r}_{0}=\alpha\,r_{0}\,,\,\,\,\,\tilde{h}=\alpha\,h
\,.\qquad
\end{eqnarray}
Subsequently, the tildes are dropped.

As for the temperature dependence we assume that the system is in
contact with a heat bath which based on molecular dynamics
techniques is simulated with the Nos\'e-Hoover method
\citep{Nose1,Nose2,Hoover}. According to this method the DNA
lattice system is coupled to two additional variables, $p_s$ and
$s$, such that the equations read as
\begin{eqnarray}
\dot{x}_{n,i}&=&p_{n,i}^{(x)}\,,\label{eq:xdot}\\
\dot{p}_{n,i}^{(x)}&=&2D_n\left[\,\exp(-d_n)-1\,\right]\,\exp(-d_n)\,
\frac{\partial d_n}{\partial x_{n,i}}\nonumber\\
&-&2K\left[\,l_{n,i}\frac{\partial l_{n,i}}{\partial x_{n,i}}+
l_{n+1,i}\frac{\partial l_{n+1,i}}{\partial
x_{n,i}}\,\right]\nonumber\\
&+&S\,[\,d_{n+1}-2\,d_{n}+d_{n-1}\,]\,\frac{\partial d_n}{\partial
x_{n,i}}\,-\,\frac{p_{n,i}^{(x)}p_s}{Q}\,,\\
\dot{y}_{n,i}&=&p_{n,i}^{(y)}\,,
\label{eq:ydot}\\
\dot{p}_{n,i}^{(y)}&=&2D_n\left[\,\exp(-d_n)-1\,\right]\,\exp(-d_n)\,
\frac{\partial d_n}{\partial y_{n,i}}\nonumber\\
&-&2K\left[\,l_{n,i}\frac{\partial l_{n,i}}{\partial y_{n,i}}+
l_{n+1,i}\frac{\partial l_{n+1,i}}{\partial
y_{n,i}}\,\right]\nonumber\\
&+&S\,[\,d_{n+1}-2\,d_{n}+d_{n-1}\,]\,\frac{\partial d_n}{\partial
y_{n,i}}\,-\,\frac{p_{n,i}^{(y)}p_s}{Q}\,, \label{eq:pydot}\\
\dot{z}_{n,i}&=&p_{n,i}^{(z)}\,,\label{eq:zdot}\\
\dot{p}_{n,i}^{(z)}&=&2D_n\left[\,\exp(-d_n)-1\,\right]\,\exp(-d_n)\,
\frac{\partial d_n}{\partial z_{n,i}}\nonumber\\
&-&2K\left[\,l_{n,i}\frac{\partial l_{n,i}}{\partial z_{n,i}}+
l_{n+1,i}\frac{\partial l_{n+1,i}}{\partial
z_{n,i}}\,\right]\nonumber\\ &+&S\,[\,d_{n+1}-2\,d_{n}+d_{n-1}\,]
\,\frac{\partial d_n}{\partial z_{n,i}}\,-\,\frac
{p_{n,i}^{(z)}p_s}{Q}\nonumber\\
&-&C\left(\,2\,z_{n,i}-z_{n+1,i}-z_{n-1,i}\,\right)\,,\label{eq:pzdot}\\
\dot{p}_s&=&\frac{1}{m}\sum_{n,i}\,\left[\,
(p_{n,i}^{(x)})^2+(p_{n,i}^{(y)})^2+(p_{n,i}^{(z)})^2\,\right]-Nk_BT\,,\nonumber\\
&&\\ \dot{s}&=&\frac{s p_s}{Q}\label{eq:sdot}\,.
\end{eqnarray}
Note that the heat bath variable $s$ does not enter explicitly the
equations of motion regarding the actual dynamics of the bases. On
the other hand $\hat{H}=H+p_s^{2}/(2Q)+Nk_BT\ln(s)$ is an
augmented Hamiltonian the conservation of which can be used to
assure accuracy of the numerical computations. The parameter $Q$
determines the time scale of the thermostat with temperature $T$
and $k_B$ is the Boltzmann constant.

The values of the scaled parameters are given by $K=0.683$,
$r_0=44.50$, $h=15.13$ and $l_0=31.39$. One time unit of the
scaled time corresponds to $0.198\,\mathrm{ps}$ of the physical
time.

%%%%%%%%%%%%%%%%%%%%%%%%figure1
\begin{figure}[t]
\centering
\includegraphics[angle=270,width=0.46\textwidth]{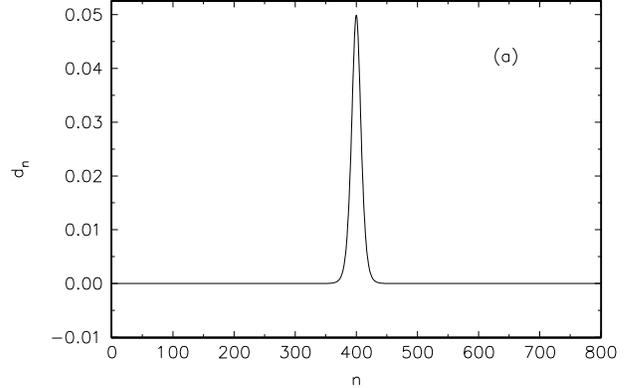}
\caption{\label{fig1} The initial radial distortions of the double
helix. The inter-base distance $d_n(0)$ in \AA.}
\end{figure}

%%%%%%%%%%%%%%%%%%%%%%%%figure2
\begin{figure}[t]
\centering
\includegraphics[angle=270,width=.45\textwidth]{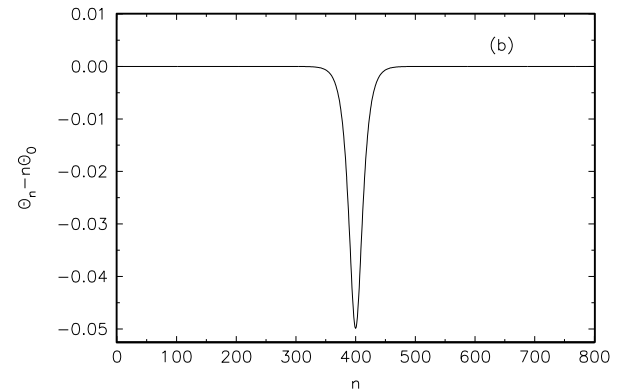}
\caption{\label{fig2} The initial angular distortions of the
double helix.}
\end{figure}

We turn now to the study of the creation of bubbles in DNA. The
starting point is a DNA molecule for which a certain segment
experiences initially angular and radial deformations due to the
action of some enzyme to which a region of the DNA is bound. In
order to simulate the deforming action of enzymes we assume that
initially a number of consecutive sites in the center of the DNA
lattice (hereafter referred to as the {\it central region}) are
exerted to forces acting in angular and radial direction such that
in this region the molecule experiences twist reduction together
with radial stretchings. These structural deformations can be
extended over a region encompassing up to thirty base pairs and as
it is going to be demonstrated give rise to the formation of
H-bridge breather solutions (extending over $15-20$ base pairs)
reproducing the oscillating 'bubbles' observed for the DNA-opening
process \citep{Barbithesis}. In Figs.~\ref{fig1} and \ref{fig2}
the localized initial distortions are shown. Both the angular and
radial deformation patterns are bell-shaped, due to the fact that
the enzymatic force is exerted locally, i.e. in the extreme case
to a single base pair only for which the H-bond is deformed
\footnote{It turns out that this shape is also more stable in a
thermal bath than a rectangular shape as in  \citealt{Hennig}.}.
The first one being of non-positive amplitudes is linked with
reduced twist while the latter one with non-negative amplitudes is
associated with radial stretchings. The radial and angular
deformation patterns are centered at the central lattice site
(base pair) at which the H-bond stretching and twist angle
reduction is at maximum. At either side of the central site the
amplitudes approach progressively zero. The deformation energy
amounts to $0.0362\,\mathrm{eV}$.
 The set of coupled equations (\ref{eq:xdot})-(\ref{eq:sdot}) is
integrated with the help of a fourth-order Runge-Kutta method. For
the simulation the DNA lattice consists of $799$ sites and open
boundary conditions were imposed. The same initial conditions are
used both at zero and nonzero temperatures.

\section{Results}
\begin{figure}[t]
\centering
\includegraphics[angle=270,width=0.46\textwidth]{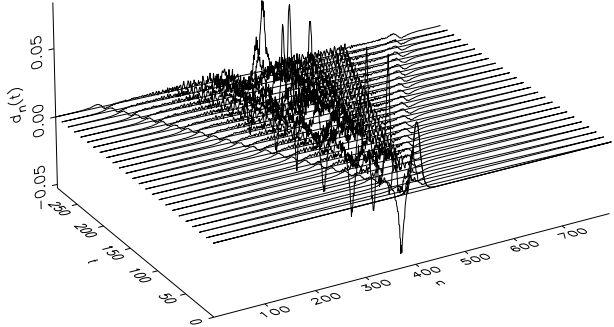}\\
\caption{\label{fig3} The case $T=0$: Spatio-temporal pattern of
the inter-base distance $d_n(t)$ in \AA. Parameters: $C=0.126$,
$K=0.63$, $S=1.26$, $D_0=1$, $D_1=2$.}
\end{figure}

First of all we consider the zero temperature regime. In
Fig.~\ref{fig3} we depict the spatio-temporal evolution of the
distance, $d_n(t)$, between two bases of a base pair, which
measures the variation of the length of the corresponding hydrogen
bond expressed in \AA. The dynamics starts from a non-equilibrium
configuration so that energy redistribution takes place. Most
noticeably, this is established in the immediate emission of
(small-amplitude) phonon waves from either side of the localized
radial pattern situated around the central lattice site. These
primary  phonon waves travel uniformly towards the ends of the
lattice with uniform velocity. In the approach of an equilibrium
regime small amount of excitation energy is dispersed in the form
of secondary phonon waves in the rest of the DNA lattice. However,
the vast majority of the excitation energy remains contained in
the central region.

%%%%%%%%%%%%%%%%%%%figure3

 We observe that the localized bell-shaped radial pattern is
preserved and performs periodic oscillations in time, viz. a
radial breather is formed. This periodically oscillating pattern
of the radial variable, $d_n(t)$, is attributed to successive
stretchings and compressings of the hydrogen bonds.

Note that the stretching of a base pair distance is larger than
the compression typical for the evolution in a Morse potential
(see also \citealt{BCP}).

\begin{figure}[t]
\centering
\includegraphics[angle=270,width=0.46\textwidth]{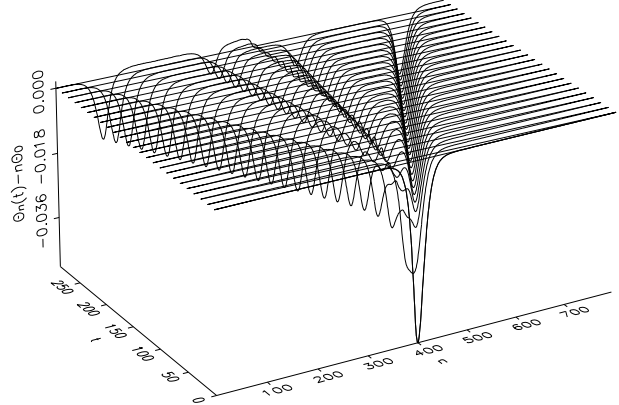}\\
\caption{\label{fig4} The case $T=0$: Spatio-temporal pattern of
the twist angle  expressed in rad. Parameters as in
Fig.~\ref{fig3} and heat bath parameter $Q=10$.}
\end{figure}

With regard to the associated pattern of the deviations of the
twist angles from their equilibrium values,
$\theta_n(t)-n\theta_0$, expressed in rad, we find that in an
initial phase the amplitudes increase. With the emanation of the
primary radial phonon waves from the localized radial pattern (cf.
Fig.~\ref{fig3})  there is linked a split up of the standing
angular bell-shaped pattern so that two primary torsional waves of
negative localized angular deformations, symmetrically to the
central site, get produced which travel in unison with the radial
phonon waves in the direction of the lattice ends. Since the
radial phonon waves possess positive amplitudes corresponding to
H-bond stretching the region of the double helix traversed by them
experiences reductions of the twist angle, i.e.
$\theta_n(t)-n\theta_0$ is negative. Similarly, the second radial
phonon waves are connected with two small negative-amplitude
angular waves following the primary ones away from the central
region. Around the central lattice site the (initially negative)
amplitudes of the twist angle grow steadily. Finally, after nearly
$150$ time units the twist angles of the helix in the breather
region become positive being equivalent to increased twist in a
region comprising 20 lattice sites to either side of the central
site. Notice that the process of amplitude breathing of the
torsional part of the lattice proceeds with a longer period than
the breathing of the radial localized pattern. This behavior of
$\theta_n(t)-n\theta_0$ for zero temperature is shown in
Fig.~\ref{fig4}.

%%%%%%%%%%%%%%%%figure4

\begin{figure}[t]
\centering
\includegraphics[angle=270,width=0.46\textwidth]{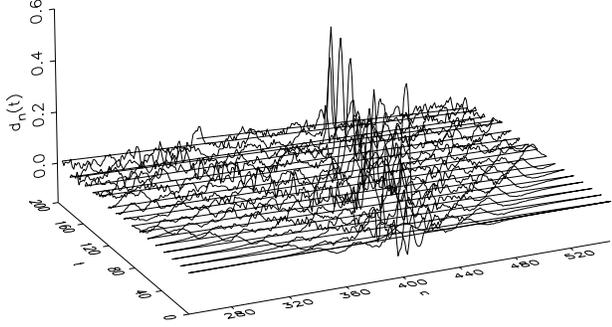}\\
\caption{\label{fig5} The case $T=100\,\mathrm{K}$:
Spatio-temporal pattern of the inter-base distance $d_n(t)$.
Parameters as in Fig.~\ref{fig3}. For better illustration only a
segment of the DNA lattice is shown.}
\end{figure}

Remarkably, in the non-zero temperature regime, when thermal
energy is injected into the DNA lattice, the radial breather
basically remains in localized shape. Moreover, its amplitudes and
energy  grow. Therefore, the breather extracts energy from the
thermal bath.
 Simultaneously we observe that excess energy, not to be carried
by the radial breather, is ejected from the central region where
it disperses into wider parts of the lattice as phonons.
Nevertheless, a radial breather of fairly large amplitude prevails
 over the small-amplitude noisy background
caused by the heat bath on the DNA lattice.
 Interestingly,
under the impact of thermal perturbations the radial breather
starts to move towards one end of the DNA lattice in coherent
fashion with maximum amplitude being larger then the one in the
$T=0$ case. This feature is illustrated for the case $T=100$ in
Fig.~\ref{fig5}. Correspondingly, the propagating radial breather
is accompanied by a torsional breather leaving the double helix
consecutively in over-twisted and under-twisted shape. The
direction of the movement of the breather is determined by the
actual realization of the random sequence of the DNA base pairs.
This is reflected in our model in bi-valued dissociation levels
$D_1 =2D_0$ that are assigned to the sites in a random fashion
along the backbone so that the translational invariance of the
lattice is broken by disorder. For the simulations presented by us
in the manuscript, that were performed with the same realization
of disorder for all considered temperatures, the breather moves
towards the right end of the DNA chain. However we found that
there exist also realizations of the random $D_0-D_1$ sequence for
which the motion heads towards the left end. The fact that the
breather movement is not diffusive, i.e., with random changes of
directions, means that the moving breather is a stable new entity
produced by the perturbation of the standing one. In mathematical
terms, the thermal bath brings about the crossing of a separatrix
between the basins of attraction of both breathers. Varying the
temperature yields qualitatively equal results, i.e. the radial
and torsional breather become eventually mobile traveling towards
one end of the DNA lattice. Generally, it holds that the higher
the temperature the more grow the amplitudes of the breather and
the stronger the helix unwinds. For ambient temperature, i.e.
$T=300\,\mathrm{K}$, the maximal amplitude of the radial breather
is ten times larger than the one of the zero-temperature case
being related to stretching of the associated hydrogen bond by
$0.6$\,\AA\, away from its equilibrium value. Moreover, at the
sites (base pairs) where the helix is radially stretched to the
most extend it is unwound by an angle of $12^{\,\circ}$.

%%%%%%%%%%%%%%%%figure5

Quantitatively, the results regarding the temperature dependence
of the breather evolution are suitably summarized by the
time-evolution of the first momentum of the energy distribution
defined as
 \begin{equation}
\bar{n}(t)=\sum_{i=1,2}\,\sum_{n=1}^{N}\,(n_c-n)\,E_{n,i}(t)
\label{eq:momentum}\,,
\end{equation}
and the  energy $E_{n,i}$ is defined in Eq.\,(\ref{eq:Eloc}) and
$n_c$ is the site index corresponding to the center of the DNA
lattice.

\begin{figure}[t]
\centering
\includegraphics[angle=270,width=0.46\textwidth]{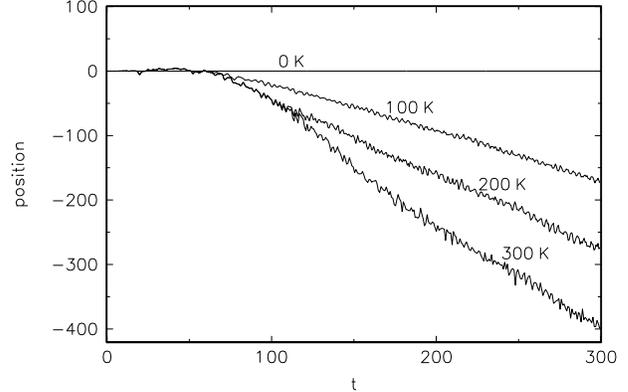}\\
\caption{\label{fig6} Influence of temperature on the
time-evolution of the breather center. Temperature as indicated on
the curves.}
\end{figure}

This quantity describes the temporal behavior of the position of
the center of a breather. Thus it represents a  measure for the
mobility of the breathers. Generally, the higher the temperature
the larger the amplitude of the radial breather becomes and the
faster the radial (and torsional) breather travels along the DNA
lattice which is illustrated in Fig.~\ref{fig6}. As the degree of
energy localization in the breathers is concerned, the energetic
partition number serves as an appropriate mean. It is defined as
\begin{equation}
P(t)=\frac{(\sum_{i=1,2}\sum_{n=1}^{N}\,E_{n,i})^2}
{\sum_{i=1,2}\sum_{n=1}^{N}\,E_{n,i}^2}\,.
\end{equation}
$P$ quantifies how many sites are excited to contribute to the
breather pattern. In the $T=0$ case the partition number changes
only slightly and performs  oscillation in an interval being
rather close
 to its initial value for all the time.
The relatively small growth of $P(t)$ accounts for the phonons
being emitted from the central region.
%%%%%%%%%%%%%%%%%%%%%figure6
%%%%%%%%%%%%%%%figure7
\begin{figure}[t]
\centering
\includegraphics[angle=270,width=0.46\textwidth]{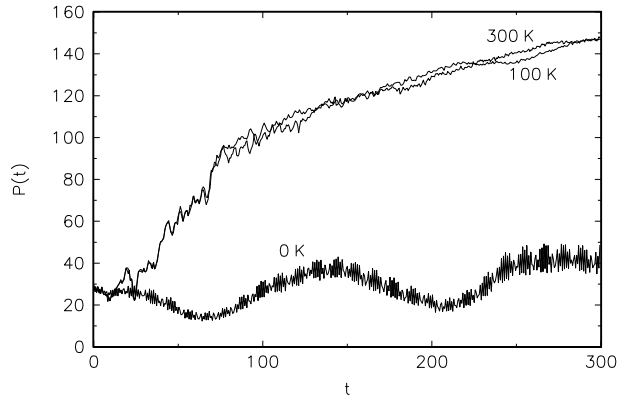}\\
\caption{\label{fig7} Temperature dependence of the temporal
behavior of the partition number. Temperature as indicated in the
graph.}
\end{figure}

For $T>0$ we find that $P(t)$ rises with time. While up to times
$t\lesssim 80$ the mean growth of $P(t)$ proceeds linearly,
afterwards, when the DNA lattice system comes more and more closer
to equilibrium,  the participation number increases only in a
logarithmic fashion. Eventually, the value of $P(t)$ reaches a
plateau around the
  time value of
 $150$. Compared to the initial width of the localized distortion
pattern, i.e. $P(0)=29$, the energy spread has risen by $\lesssim
44\%$. This energy distribution over the lattice, however, is
mainly due to the thermalization of the lattice by the thermal
perturbation rather then by an actual spread of the radial
breather. The latter maintains its degree of localization
throughout the time.
  The time evolution of $P$ is shown in
Fig.~\ref{fig7}.
 Remarkably, in the range of $50\,\mathrm{K}\leqslant300\,\mathrm{K}$ we
observe almost equal temporal evolution of $P$ regardless of the
value of $T$. Thus the thermal perturbations do not strongly
influence the width of the breather anymore giving evidence of the
stability of the breather and the pronounced storing capability of
the DNA lattice

\section{Discussion}
In conclusion, we observed that out of an initial non-equilibrium
situation, for which the hydrogen bonds of an under-twisted
segment of the DNA lattice have been stretched, breathers develop
in the radial and angular displacement variables. For zero
temperature the breathers remain standing in the initially excited
region. Interestingly, for $T>0$ the breathers start to travel
coherently towards an end of the DNA duplex. The observed
breathers represent realistically the oscillating bubbles found
prior to complete unzipping of DNA, i.e. they oscillate with
periods in the range $0.3\,-\,0.8\,ps$, are on the average
extended over $10-20$ base pairs and possess maximal amplitudes of
the order of $\lesssim 0.6$\,\AA. Our results demonstrate that the
action of some enzyme, mimicked by localized radial and torsional
distortions of the DNA equilibrium configuration, initiates in
fact the production of oscillating bubbles in DNA. Moreover, these
oscillating bubbles sustain the impact of thermal perturbations.
Whether the bubbles, when meeting each other, can merge such that
a larger-amplitude radial breather results which resembles the
replication bubble with broken H-bonds is still being
investigated.

\vspace{0.5cm} \centerline{\large{\bf Acknowledgments}}

\noindent The author acknowledges  support by the Deutsche
Forschungsgemeinschaft via a Heisenberg fellowship (He 3049/1-1).
JM Romero acknowledges a Study License from the Junta de
Andalucia, Spain.

%%%%%%%%%%%%%%%%%%%%%%%%%%%%%%%%%%%%%%%%%%%%%%%%%%%%%%%%%%%%%%%%%%%%%%%%%%%%%
\newcommand{\noopsort}[1]{} \newcommand{\printfirst}[2]{#1}
  \newcommand{\singleletter}[1]{#1} \newcommand{\switchargs}[2]{#2#1}

%%%%%%%%%%%%%%%%%%%%%%%%%%%%%%%%%%%%%%%%%%%%%%%%%%%%%%%%%%%%%%%%%%%%%%%%%%%%55
\end{document}